\documentclass[12 pt]{article}
\usepackage[left=1in, right=1in, top=0.75in, bottom=0.75in]{geometry}
\usepackage[utf8]{inputenc}
\usepackage{textcomp}
\usepackage{setspace}
\usepackage{graphicx}
\usepackage{chemformula}
\usepackage{authblk}
\usepackage{booktabs}
\usepackage{gensymb}
\usepackage{amsmath}
\usepackage{parskip}
\usepackage{caption}
\usepackage{titlesec}
\usepackage{wrapfig}

\usepackage[bookmarks=false,colorlinks,citecolor=blue]{hyperref}
\hypersetup{colorlinks=true,linkcolor=blue,filecolor=blue,citecolor = blue, urlcolor=blue}

\titlespacing{\section}{0pt}{12pt plus 2pt minus 2pt}{12pt plus 2pt minus 2pt}

\usepackage{sectsty}
\sectionfont{\fontsize{14}{16}\selectfont}
\subsectionfont{\fontsize{12}{13}\selectfont}

\usepackage[backend=biber, style=chem-acs,articletitle=true,sorting=none]{biblatex}
\newcommand{\CrxNbS}{\ch{Cr_{1/3}NbS2}}
\newcommand{\CrxTaS}{\ch{Cr_{1/3}TaS2}}

\newcommand{\CoxMS}{\ch{Co_{1/3}\emph{M}S2} 
(\emph{M} = Nb, Ta)}

\newcommand{\Hc}{$H_\textrm{c}$}
\newcommand{\Tc}{$T_\textrm{C}$}

\title{\Large\textbf{Consequences and control of multi-scale (dis)order in chiral magnetic textures}}

\author[1,$\dagger$]{Berit H. Goodge}
\author[1,$\dagger$]{Oscar Gonzalez}
\author[1]{Lilia S. Xie}
\author[1,2,*]{D. Kwabena Bediako}
\affil[1]{Department of Chemistry, University of California, Berkeley, CA 94720, USA}
\affil[2]{Chemical Sciences Division, Lawrence Berkeley National Laboratory, Berkeley, CA 94720, USA}
\affil[*]{Correspondence to: bediako@berkeley.edu}
\affil[$\dagger$]{These authors contributed equally to this work}

\date{}
\begin{document}
\maketitle

\doublespacing

\section*{Abstract}

Transition metal-intercalated transition metal dichalcogenides (TMDs) are promising platforms for next-generation spintronic devices based on their wide range of electronic and magnetic phases, which can be tuned by varying the host lattice or the identity of the intercalant, along with its stoichiometry and spatial order.
Some of these compounds host a chiral magnetic phase in which the helical winding of magnetic moments propagates along a high-symmetry crystalline axis. 
Previous studies have demonstrated that variation in intercalant concentrations can have a dramatic impact on the formation of chiral domains and ensemble magnetic properties. 
However, a systematic and comprehensive study of how atomic-scale order and disorder impacts collective magnetic behavior are so far lacking. 
Here, we leverage a combination of imaging modes in the (scanning) transmission electron microscope (S/TEM) to directly probe (dis)order across multiple length scales and show how subtle changes in the atomic lattice can be leveraged to tune the mesoscale spin textures and bulk magnetic response, with direct implications for the fundamental understanding and technological implementation of such compounds. 

\newpage

\section*{Introduction}

Future electronic devices may rely on the manipulation of spin for information storage, mandating the exploration of solid-state platforms that enable magnetic order to be finely tuned and controlled.\supercite{wolf_spintronics_2001,fert2013,parkin2015,tey2022} 
The potential benefits of miniaturization in terms of storage density and/or power efficiency may be realized either through the design of magnetic materials in which the atomic lattice imposes nanoscale confinement (that is, low-dimensional magnetic materials),\supercite{Burch2018,Gong2019} or by exploiting atomic lattices—even in bulk three-dimensional materials—which produce nanoscale spin textures owing to a balance of disparate spin–spin correlations.\supercite{Yang2021,Song2021,Muhlbauer2009} 
Transition metal-intercalated transition metal dichalcogenides (TMDs) offer a rich platform to investigate a wide range of magnetic phenomena.\supercite{Eibschutz1981,volkova2014role,maniv_exchange_2021, morosan_sharp_2007, Choi2009,Hardy2015, husremović2022hard, park_field-tunable_2022, zhang_chiral_2021, togawa2012chiral, hall2021magnetic,mayoh2022giant, edwards2023giant} 
These materials can be described by the general chemical formula $T_xMCh_2$, where $T$ and $M$ are transition metals, $Ch$ is a chalcogen, and $x<$1. 
The intercalant stoichiometry $x$ can direct the formation of superlattices, through long-range ordering of the intercalant ions. 
These intercalation-derived superlattice structures in turn alter the overall symmetry of the crystal and dictate the bulk magnetic properties.\supercite{xie2022structure} 
For example, first-row transition metal intercalants have been found to result in $2a_0\times2a_0$ (principally at $x$ = 1/4) or $\sqrt{3}a_0\times\sqrt{3}a_0$ (principally at $x$ = 1/3) superlattices (where $a_0$ is the in-plane lattice constant of the non-intercalated $MCh_2$ host lattice) when the intercalants occupy the pseudo-octahedral sites between $MCh_2$ layers.\supercite{xie2022structure} 
The $\sqrt{3}a_0\times\sqrt{3}a_0$ superlattice in particular introduces noncentrosymmetry and chirality into the structure, giving rise to antisymmetric exchange interactions (also referred to as Dzyaloshinskii--Moriya, or DM, interactions)\supercite{dzyaloshinsky1958,moriya_anisotropic_1960} that favor the out-of-plane canting of spins and compete with ferromagnetic exchange.

\begin{figure}[htbp]
    \centering
        \includegraphics[width=\linewidth]{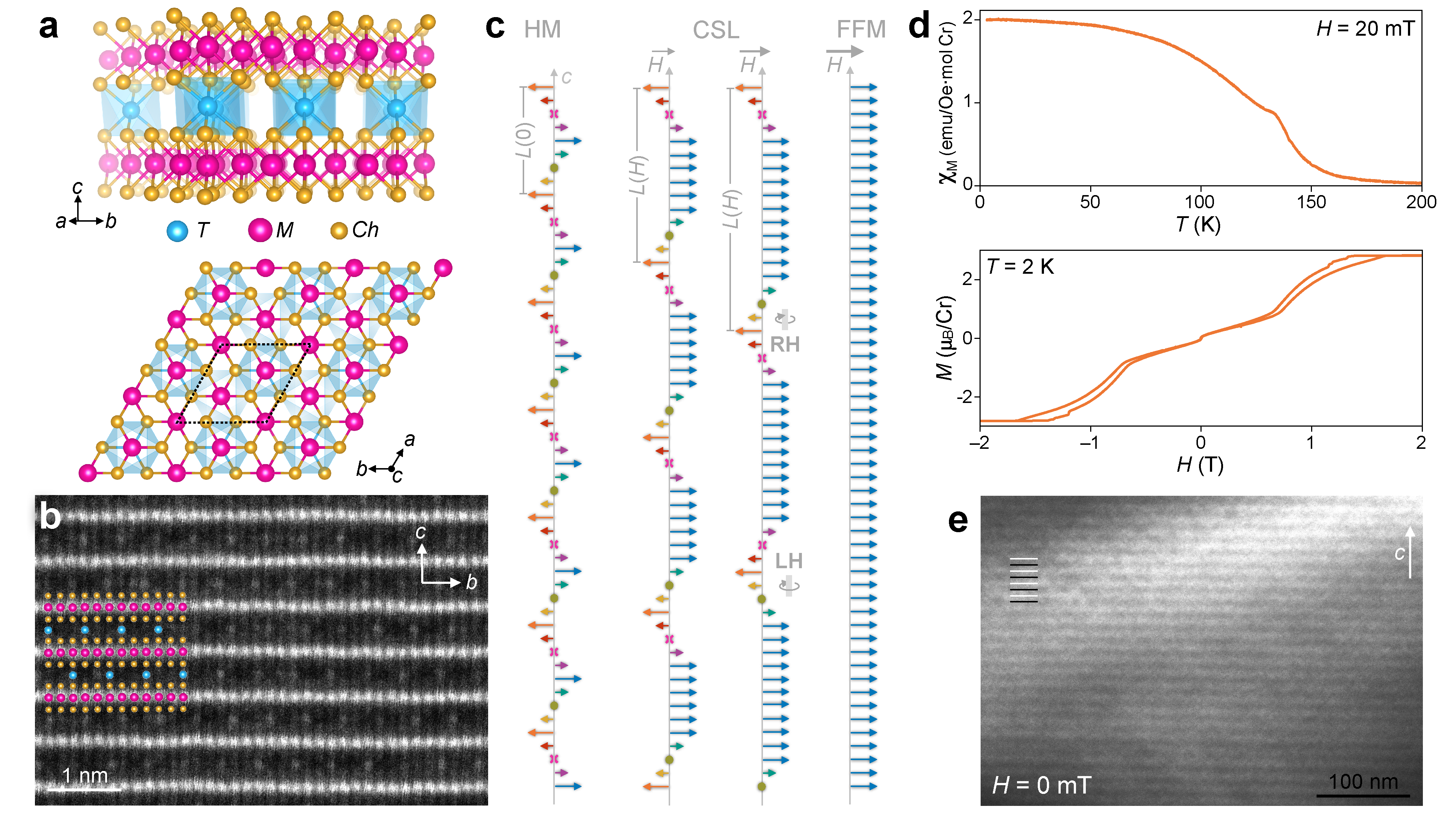}
    \caption{\textbf{Tunable chiral magnetic textures in intercalated transition metal dichalcogenide (TMD) compounds.} \textbf{(a)} Two views of the crystal structure of $T_{1/3}MCh_2$ where $T$ is the intercalant, $M$ is the metal ion, and $X$ is the chalcogen. The dashed parallelogram indicates the $\sqrt{3}a_0\times\sqrt{3}a_0$ supercell where $a_0$ is the lattice constant of the host $MCh_2$ lattice. \textbf{(b)} Atomic-resolution HAADF-STEM image and schematic overlay of \CrxTaS, (Cr = cyan, Ta = magenta, S = gold). \textbf{(c)} Illustration of magnetic order along the crystallographic $c$-axis with spins in each plane oriented uniformly but with an angle offset with respect to adjacent layers. The zero-field helimagnet (HM) phase has a characteristic period $L(0)$. With increasing applied fields, $\vec{H}$, the magnetic spiral begins to ``unwind,'' leading to chiral soliton lattice (CSL) phases with increasing periodicities $L(H)$ and, eventually, to a forced ferromagnetic (FFM) state. Soliton walls can follow left- or right-handed winding as pictured in the second CSL phase. \textbf{(d)} Representative bulk magnetic measurements of \CrxTaS, showing the magnetic susceptibility, $\chi{_M}$, as a function of temperature, $T$, (top) and the magnetization, $M$, as a function of magnetic field, $H$ (bottom). The field was applied in the \textit{ab}-plane.  \textbf{(e)} Zero-field cryo-LTEM image showing the chiral helimagnetic order in \CrxTaS, with a $L(0) \sim 17$ nm.  \label{fig:intro}}
\end{figure}

The intercalation of Cr into niobium and tantalum disulfides to produce \CrxNbS\ and \CrxTaS\ compounds leads to structures that possess this $\sqrt{3}a_0\times\sqrt{3}a_0$ superlattice, as depicted in Figure \ref{fig:intro}a and Supplemental Figure S1.\supercite{rouxel1971} 
In Figure \ref{fig:intro}b, the ordered arrangement of Cr intercalants between 2$H$-TaS$_2$ layers is directly observed through atomic-resolution high-angle annular dark-field scanning transmission electron microscopy (HAADF-STEM) where the image contrast depends on elastic scattering of primary electrons in the STEM probe such that heavier nuclei give rise to brighter contrast. 
This electron micrograph shows how the Cr sublattices in each intercalant layer are offset along [010] from those in the adjacent layers such that the intercalants do not occupy the interstitial sites directly above or below each other.

The broken inversion symmetry that arises from these intercalant positions and attendant DM interactions result in chiral helimagnetic (HM) phases\supercite{Miyadai1983,togawa2012chiral,togawa_symmetry_2016,aczel2018,xie2022structure} that evolve from a zero-field texture comprising layers of aligned spins rotated relative to each other along the direction of the helical winding, which is the crystallographic $c$-axis (Figure \ref{fig:intro}c).
Under an external magnetic field applied in the easy ($ab$) plane, these chiral textures begin to ``unwind'' such that regions of locally aligned spins are periodically separated by magnetic solitons, forming what is known as a chiral soliton lattice (CSL).\supercite{Miyadai1983, moriya_evidence_1982,chapman2014spin}
Above a critical field strength, \Hc, all the spins align along the direction of the external field in a so-called forced ferromagnetic (FFM) state.\supercite{moriya_evidence_1982}

This magnetic evolution can be observed in measurements of magnetic susceptibility as a function of temperature and magnetization as a function of applied magnetic field, which express a kink around the Curie temperature, \Tc, and a small hysteresis around \Hc, respectively (Figure \ref{fig:intro}d), both of which have been attributed to CSL formation and evolution.\supercite{Miyadai1983, moriya_evidence_1982} 
In addition, the HM texture can be directly imaged with Lorentz transmission electron microscopy (LTEM), as shown in Figure \ref{fig:intro}e for the zero-field HM phase in $\CrxTaS$. 
In this measurement, the deflection of high-energy primary electrons in the TEM beam due to local moments in the sample creates dark and light contrast in over- and under-focus imaging conditions.\supercite{chapman1984investigation, togawa2012chiral} 
Consequently, in Figure \ref{fig:intro}e, the periodic horizontal stripes of bright and dark lines arise from the one-dimensional winding magnetic order (Supplemental Figure S2), revealing a 17 nm HM periodicity, $L(0)$, consistent with previous reports.\supercite{du2021topological}

Previous work within the family of intercalated TMD compounds has investigated mesoscale evolution of magnetic order\supercite{togawa2012chiral, togawa2015magnetic,du2021topological} and the subtle dependence of bulk magnetization response on a variety of tuning knobs, including stoichiometry,\supercite{horibe2014color,kousaka2022emergence, wu2022highly} mechanical strain,\supercite{paterson2020tensile} and sample geometry.\supercite{li2023transformation}
So far, however, a comprehensive study uniting structural and magnetic characterization across length scales from the atomic- to micro-scale has been lacking. 
Such an analysis is critical for these materials because, as intercalation compounds, they are distinctively susceptible and sensitive to disorder on the intercalant (magnetic) sublattice, with severe consequences for the emergent magnetic properties.\supercite{xie2022structure,morosan_sharp_2007,dyadkin2015structural,Hardy2015,maniv_exchange_2021}

Here, we probe the structure and consequent magnetic behavior of the HM systems, with a primary focus on Cr-intercalated $2H$-\ch{NbS2}, \CrxNbS. 
We interrogate the impacts of atomic order and disorder on mesoscopic magnetic textures and bulk properties of the emergent HM phases and CSL transitions. 
We show that subtle, local stoichiometric variations can lead to pronounced (up to three-fold) changes to the CSL periodicity. 
Atomic-resolution electron microscopy unveils how these quantitative distinctions arise from different mechanisms for accommodating minimal extents of Cr deficiency, which depend on the rate at which crystals are cooled during synthesis.
Beyond magnetic periodicity, we find that concentrations of disorder in the atomic lattice additionally nucleate mesoscopic defects in the magnetic lattice, including dislocations, shearing, and heterochirality. 
Our study reveals the propagating impact of local and global order and disorder in intercalated TMD compounds and identifies key parameters for engineering high-quality, predictable materials for future fundamental studies and technological applications.

\section*{Results and Discussion}
\subsection*{Structure and chiral helimagnetic texture in \CrxNbS}

Single crystals of \CrxNbS\ were synthesized from the elements by chemical vapor transport at growth temperatures of 950–1000 $\degree$C, as detailed in the Methods. 
Two batches of crystals, differing principally in the cooling rate, were grown: one batch was cooled from the growth temperature at a rate of 20 $\degree$C/hr, whereas the second batch was cooled more rapidly at 60 $\degree$C/hr. 
Crucially, notwithstanding these synthetic variations, both batches of \CrxNbS\ samples are stoichiometrically identical within the error of our analysis (Supplemental Figures S3 and S4, Supplemental Table S1), suggesting that the Cr filling should vary by less than 1\% between the two crystals. 

\begin{figure} [htbp]
    \centering
        \includegraphics[width=0.63\linewidth]{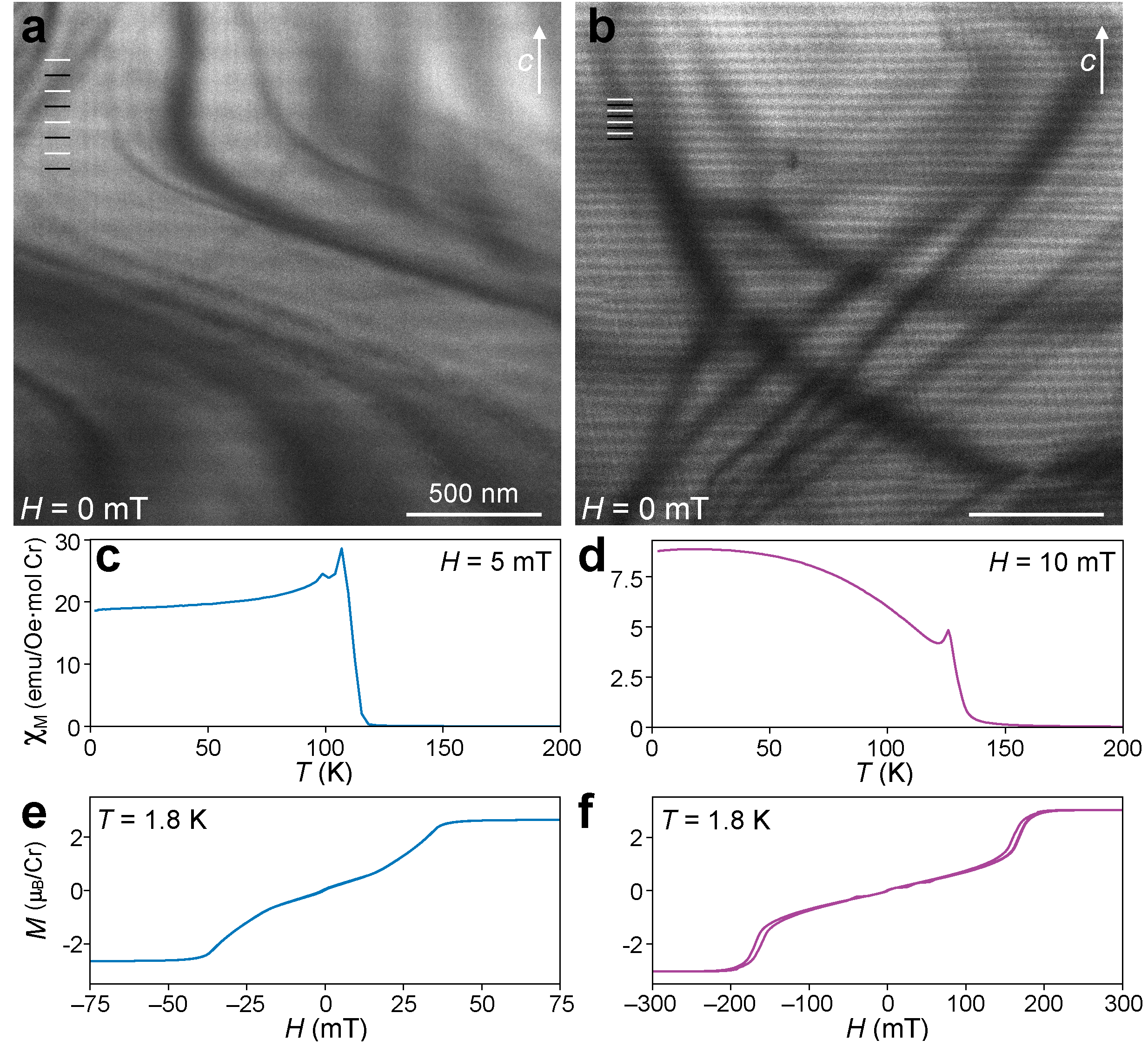}
    \caption{\textbf{Varying magnetic order in two \CrxNbS}. \textbf{(a,b)} Zero-field cryo-LTEM images showing chiral helimagnet ordering in \CrxNbS\ samples with \textbf{a, }HM $L(0) \sim 113$ nm and \textbf{b, }HM $L(0) \sim 43$ nm. \textbf{(c,d)} Magnetic susceptibility ($\chi{_M}$) as a function of temperature, $T$, for the two \CrxNbS\ samples in \textbf{a} and \textbf{b}, respectively. \textbf{(e,f)} Magnetization, $M$, as a function of magnetic field, $H$, for \textbf{e,} the \CrxNbS\ sample in \textbf{a} and \textbf{f,} a \CrxNbS\ sample synthesized with the same conditions as that in \textbf{b} and \textbf{d}. Comparative magnetization data for both crystals are shown in Supplemental Figure S5. The external field was applied in the \textit{ab}-plane. The static field or temperature of each measurement in \textbf{c-f} is provided in each. \label{fig:Nb-LTEM}}
\end{figure}

Figures \ref{fig:Nb-LTEM}a,b show cryo-LTEM images of cross-sectional TEM samples from these batches, revealing dramatically different HM periodicities: the slower-cooled sample exhibits a HM periodicity of $\sim$113 nm (Fig. \ref{fig:Nb-LTEM}a) and the more rapidly cooled crystal possesses a HM periodicity of $\sim$43 nm (Fig. \ref{fig:Nb-LTEM}b). 
These periodicities are respectively longer than and in good agreement with previous reports in the same compound \supercite{togawa2012chiral, li2023transformation}. 
For consistency, we refer to these as the ``long-period'' and ``short-period'' samples for the rest of our discussion. 
The single crystals from which these TEM samples were excised exhibit qualitatively similar magnetization responses with the characteristic CSL kink near \Tc\ in the $\chi_{M}(T)$ response, while the $M(H)$ behavior exhibits a low field linear region with a sharp increase in $M$ followed by saturation at \Hc, along with a negligible hysteresis (Figure \ref{fig:Nb-LTEM}c--d).
Quantitatively, however, the critical fields differ by a factor of nearly 4: the long-period sample saturates near 50 mT while the short-period sample saturates at approximately 200 mT.

\begin{figure}
    \centering
        \includegraphics[width=\linewidth]{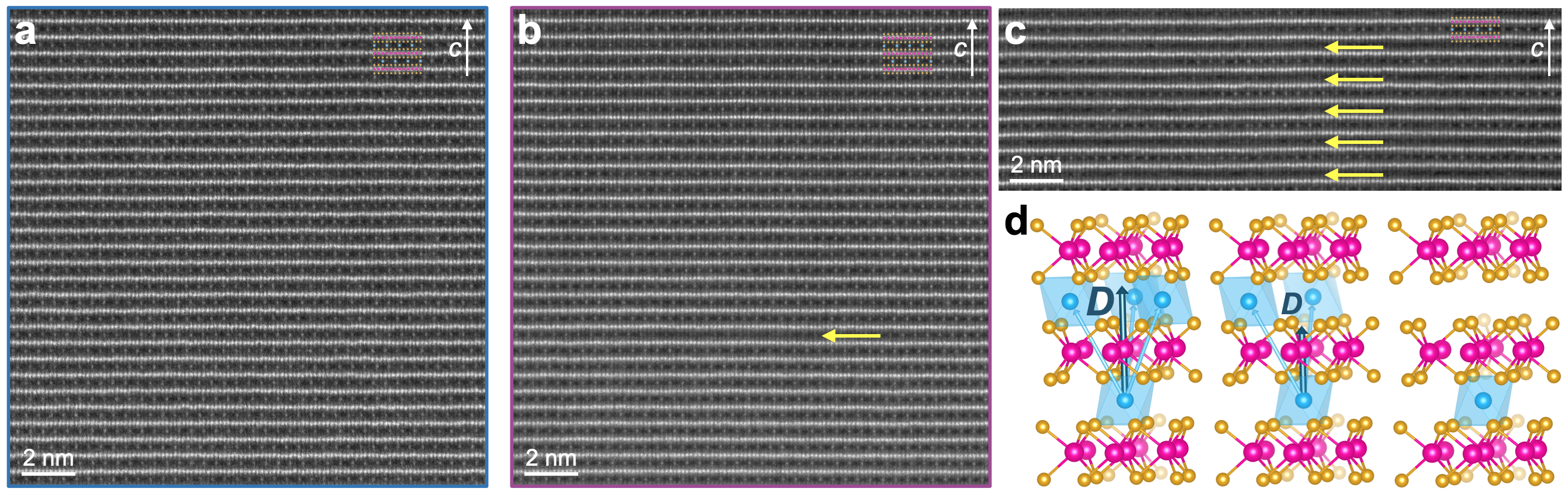}
    \caption{\textbf{Atomic-scale insights to global vs. local distribution of Cr vacancies}. \textbf{(a,b)} Atomic-resolution HAADF-STEM images of the \CrxNbS\ samples in Figure \ref{fig:Nb-LTEM}a and \ref{fig:Nb-LTEM}b, respectively, with model overlays as shown in Figure \ref{fig:intro}. A local cluster of Cr vacancies is visible in \textbf{b}, marked with a yellow arrow. \textbf{(c)} HAADF-STEM image of a Cr-deficient (Cr$_{0.28}$NbS$_{2}$) sample prepared in the same conditions as that in \textbf{b} shows a high density of Cr vacancies similarly clustered. \textbf{(d)} Schematic representation of the impact of Cr vacancies on the antisymmetric DM interaction vector $\vec{D}$ (dark teal arrow) based on addition of interlayer Cr--Cr couplings (cyan arrows) according to Moriya's rules.  \label{fig:Nd-HAADF}}
\end{figure}

To investigate the origins of these dramatic differences in HM texture of the long- and short-period samples, we leverage atomic-resolution HAADF-STEM imaging (Figures \ref{fig:Nd-HAADF}a, b). 
Both samples show overall high crystallinity, with no observable stacking faults or other defects in the parent TMD lattice, and have a mostly uniform, well-ordered distribution of the Cr intercalants consistent with the model in Figure \ref{fig:intro}a.
The short-period sample, however, exhibits local ``gaps'' in the intercalant lattice, as marked by a yellow arrow in Figure \ref{fig:Nd-HAADF}b, which are not found in the long-period sample. 
Although the projection nature of STEM imaging, which collapses a specimen foil of finite thickness into a single two-dimensional image, makes it extremely difficult to resolve point defects such as single-site vacancies, the lack of atomic contrast within the indicated few-nanometer region indicates a local concentration of Cr vacancies (an absence of Cr intercalants). 
This kind of vacancy clustering is found in more strongly deficient Cr$_{0.28}$NbS$_2$ (Figure \ref{fig:Nd-HAADF}c) as well as the related Fe-intercalated NbS$_2$ compound,\supercite{maniv_exchange_2021} and is reminiscent of the staging behavior that is common to graphite intercalation compounds.\supercite{Shu1993,Ohzuku1993,Guo2016}

As discussed above, these samples are stoichiometrically identical within the limits of our compositional analysis. 
Accordingly, we assume that both samples are equally very slightly Cr-deficient such that comparable amounts Cr vacancies are present in both samples. 
We therefore posit that an equivalent number of vacancies which are observed to form clustered pockets in the short-period sample are distributed randomly and uniformly throughout the long-period sample (where, as previously noted, they are effectively undetectable by HAADF-STEM imaging). 
We ascribe this difference in vacancy distribution to the rate at which the crystals were cooled from the elevated synthesis temperatures -- 20 and 60 $\degree$C/hr for the long- and short-period samples, respectively -- surmising that slower cooling results in a globally homogeneous distribution of vacancies, while more rapid cooling results in inhomogeneous vacancy concentrations at a local level.

This picture of Cr vacancy accommodation can further explain the vastly different magnetic periods. 
The winding of spins in the HM phase is dependent on the ratio of the DM vector between neighboring intercalants $i$ and $j$ in consecutive planes, $\vec{D}_{ij}$, and their magnetic exchange coupling, $\vec{J}_{ij}$.
The wavevector of the winding period is given by $ Q_0 = 2 \pi [L(0)]^{-1} =\arctan(D_z/J)$ \supercite{kishine2014topological}, which for small values of $D_z/J$ can be approximated and rewritten as:

\begin{equation}
\label{eq:HM period}
L(0) \approx 2\pi c \frac{|J|}{|D_z|} 
\end{equation}
where $L(0)$ is the period in the (zero-field) HM phase, $\textit{c}$ is the lattice constant along the helical axis, $J$ is the magnetic exchange coupling, and $D_z$ is the magnitude of the DM interaction.\supercite{togawa2012chiral, kishine2014topological, du2021topological} 

In compounds with similar lattice constants and identical intercalant species, the magnetic pitch can be tuned primarily through the relative strength of ${D}_{z}$,\supercite{du2021topological} where a stronger DM interaction leads to larger azimuthal misorientation between interlayer spins and a smaller period. 
One factor that determines $\vec{D}_{ij}$ is spin--orbit coupling (SOC).
In the context of $T_xMCh_2$ systems, host lattices of heavier atoms possess stronger DM interactions,\supercite{moriya_anisotropic_1960} as reflected in the smaller HM period of \CrxTaS\ (Figure \ref{fig:intro}e) than in either sample of \CrxNbS. 
Here, however, we can estimate a relative decrease in the magnitude of $D_z$ based on the expanded $L(0)$ of the long-period sample suggesting that $D_{z \textrm{, short}} \approx 2.6 \times D_{z \textrm{, long}}$ for the same host lattice (that is, same SOC and same $c$).

A schematic representation of the DM interactions that give rise to HM behavior are presented in Figure \ref{fig:Nd-HAADF}d, where couplings between a Cr ion in the bottom layer to three neighboring ions in the top layer are represented by cyan vectors between atomic sites. 
The sum of the corresponding $\vec{D}_{ij} (i\neq j)$ vectors results in an overall $\vec{D}$ which points along the crystalline $c$-axis.
When several Cr ions are missing, as in the case of a vacancy cluster, the interlayer coupling is locally fully suppressed.
On the other hand, when a single Cr ion is removed, as in the case of an isolated vacancy, the magnitude of the $\vec{D}$ vector is decreased while the overall direction remains along the crystallographic $c$-axis when averaged over random vacancies across the whole sample. 
The strength of this effect (decrease in $D_z$ by $\sim$2.6 times) for such a seemingly subtle difference in the distribution of diffuse vacancies (again, we emphasize that both samples are measured to be within 1\% of 1/3 stoichiometric Cr filling) points to an extremely sensitive dependence of the magnetic coupling in these materials on dilute interactions. 

We can thus understand the helimagnetic ordering of the short- and long-period samples by considering the localized versus distributed nature of Cr vacancies: 
the long-period sample consists of random Cr vacancies at a global scale which reduces the overall strength of $D_z$, while the short-period sample consists of locally clustered Cr vacancies such that the global magnitude of $D_z$ remains relatively large -- except at the local vacancy clusters -- and the global HM period is correspondingly shorter.
A similar concept of correlated disorder in the Cr sublattice has been previously proposed to explain variations in reported \Tc\ throughout the literature based on diffuse scattering measurements and Monte Carlo simulations,\supercite{dyadkin2015structural} but the precise atomic-scale structure of the Cr disorder has not been directly observed until now.

It is further interesting to note differences in the characteristic intercalant (dis)order in various $T_xMCh_2$ compounds: 
in seemingly stoichiometric \ch{Fe_{0.25}TaS2}, for example, changes to synthetic conditions (cooling rate) manifest as large changes to the ferromagnetic coercivity, and are attributed to variations in the population of $2a_0\times2a_0$ and $\sqrt{3}a_0\times\sqrt{3}a_0$ superlattices.\supercite{Choi2009}
On the other hand, here (and indeed, even in the more strongly deficient \ch{Cr_{0.28}NbS2} shown in Figure \ref{fig:Nd-HAADF}c) we find no signatures of mixed $2a_0\times2a_0$ ordering in addition to the $\sqrt{3}a_0\times\sqrt{3}a_0$. 
Future studies which consider systematic variation of intercalant species, stoichiometry, and host lattice may uncover if key ingredients (such as certain magnetic behavior) favor specific structural motifs. 

\subsection*{Evolution of chiral spin textures in external magnetic fields}
The technological promise of these and other magnetically textured materials lies not just in controlling their static ordering, but in dynamically and predictably tuning magnetic order with an external stimulus. 
It is therefore imperative to additionally understand the evolution of the magnetic textures under an applied field. 
Here, we find that the two \CrxNbS\ samples undergo remarkably different CSL phase evolution, which further emphasizes the impact of local versus global disorder in these systems. 

\begin{figure}
    \centering
        \includegraphics[width=0.75\linewidth]{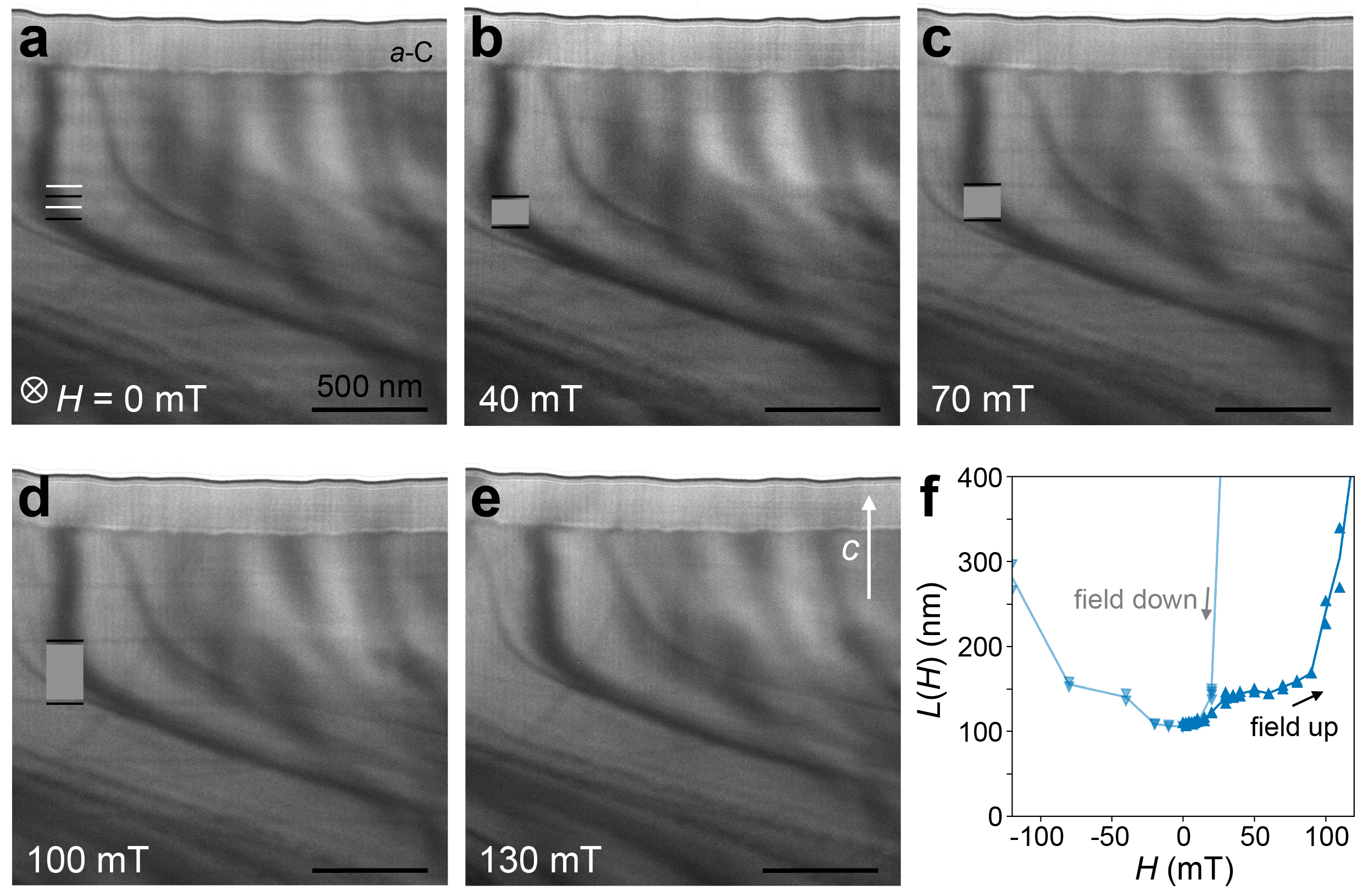}
    \caption{\textbf{Evolution of magnetic texture in the long-period \CrxNbS\ sample.} \textbf{(a–e) } Cryo-LTEM images of the chiral magnetic texture in the long-period \CrxNbS\ sample under applied magnetic field orthogonal to the $c$-axis with field strengths indicated for each frame. Light and dark lines in \textbf{a} are added to highlight the peaks and troughs of periodic HM contrast. In \textbf{b-d} the dark lines denote soliton walls and the medium gray lines mark regions of local FM ordering in the CSL phase.  \textbf{(e)} Progression of the soliton period $L(H)$ tracked by \emph{in situ} LTEM imaging under increasing (dark, upwards-pointing triangles) and decreasing (light, downwards-pointing triangles) magnetic fields. A layer of protective amorphous carbon ($a$-C) from the specimen preparation is visible along the top of the sample; the top of each image is vacuum.  \label{fig:LTEM-slow}}
\end{figure}

Figure \ref{fig:LTEM-slow} shows the evolution of the CSL phase in the long-period \CrxNbS\ sample under an \emph{in situ} applied magnetic field. 
As expected, the soliton lattice spacing increases with the field strength until no solitons remain visible and the sample can be considered in the FFM phase above $\sim$130 mT (Figure \ref{fig:LTEM-slow}e).
The external field at which no soliton walls are visible by cryo-LTEM is somewhat larger than the critical field \Hc\ at which the magnetization $M$ appears to saturate (Figure \ref{fig:Nb-LTEM}c): $\sim$136 versus $\sim$50 mT, respectively. 
One possible explanation for this apparent discrepancy may arise from the relative sensitivities of the two measurements to very low densities of soliton walls: real-space imaging by cryo-LTEM can clearly resolve single solitons whenever the spacing is less than the dimension of the prepared lamella, here $\sim$3-4 $\mu$m. 
The magnetic volume fraction, however, contributed by solitons becomes vanishing small at very large $L(H)$ such that the bulk magnetic response may appear effectively saturated even if some small density of soliton walls remain.
To our knowledge, similarly correlated measurements by bulk and real-space techniques have not been reported for other crystals, but reports of critical fields measured by LTEM \supercite{togawa2015magnetic} and bulk magnetization \supercite{togawa2016symmetry} on samples from the same group show a similar qualitative trend.

The field-dependent CSL period $L(H)$ increases in discrete steps with abrupt changes, as has been observed previously \supercite{togawa2012chiral, togawa2015magnetic} (Figure \ref{fig:LTEM-slow}f and Supplemental Video 1).
The value of $L(H)$ shows some tendency to favor certain ``metastable'' lengths, for example remaining relatively stable near 150 nm for $H$ between 30 and 80 mT (Figure \ref{fig:LTEM-slow}f). 
We note that the CSL lengths within this plateau appear to coincide with the approximate thickness of our TEM specimen (see Supplemental Figure S6), and speculate that this may point to the relevance of some dimensional confinement or other boundary conditions as has been recently discussed in the context of fabricated nanowedges of the same material. \supercite{li2023transformation}
Future efforts to explore these effects more systematically with both experimental measurements and theoretical models should provide more insight in this regard. 

We also observe a large hysteresis in $L(H)$ when the external field is ramped down,\supercite{togawa2015magnetic} showing a much more abrupt transition out of the FFM phase with a sudden onset of soliton walls (Supplemental Figure S7).
In addition, throughout the CSL phase evolution in a given field direction (i.e. positive $H$), all of the solitons show the same chirality, as indicated by consistent dark contrast in the underfocussed Fresnel cryo-LTEM conditions. 
The soliton contrast inverts when the applied field is reversed (negative $H$), further consistent with homochirality throughout the sample and under both field directions (Supplemental Figure S8).
Overall, the magnetic phases in the long-period sample are found to be fairly well-ordered and reproducible. 

\begin{figure}
    \centering
        \includegraphics[width=0.75\linewidth]{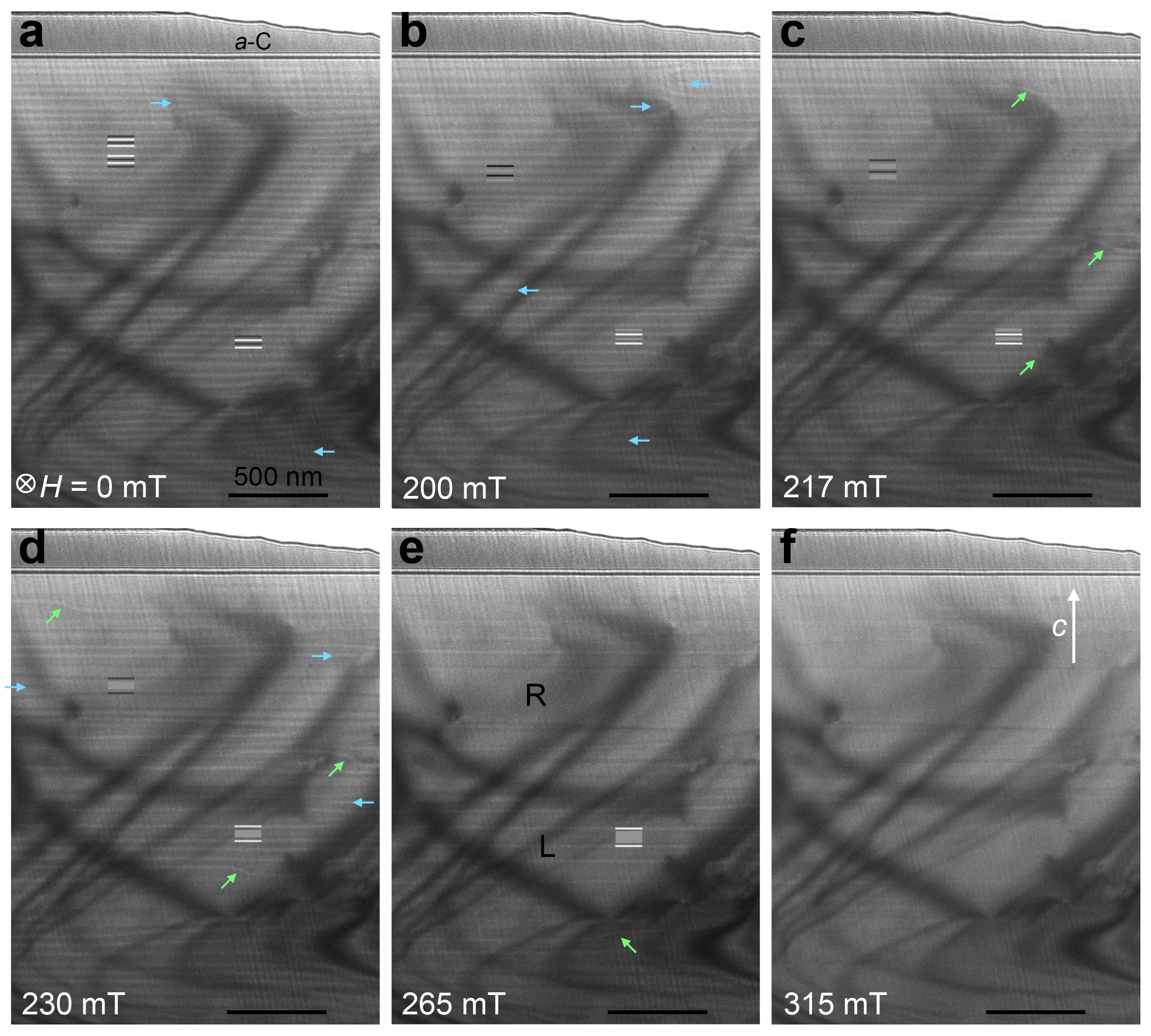}
    \caption{Evolution of magnetic texture in short-period \CrxNbS. \textbf{(a-f)} Cryo-LTEM images of the chiral magnetic texture in the short-period \CrxNbS\ sample under applied magnetic fields orthogonal to the $c$-axis with strengths indicated for each frame. Light and dark lines in \textbf{a} are added to highlight the peaks and troughs of periodic HM contrast. In \textbf{(b-f)} the dark and bright lines denote soliton walls of opposite chirality and the medium gray lines mark regions of local FM ordering in the CSL phase. Dislocations and shearing in the magnetic lattice are highlighted with blue and green arrows, respectively. A layer of protective amorphous carbon ($a$-C) from the specimen preparation is visible along the top of each micrograph; the top right corner in each image is vacuum. \label{fig:LTEM-fast}}
\end{figure}

In contrast, the magnetic texture of the short-period sample exhibits considerably different behavior (Supplemental Video 2).
Figure \ref{fig:LTEM-fast} contains a similar \emph{in situ} magnetic field cryo-LTEM image series through the CSL phase.
Unlike the long-period \CrxNbS, which progresses through uniform $L(H)$ at different fields, the short-period \CrxNbS\ exhibits widely-varied soliton spacings at every field such that a single $L(H)$ can not be reasonably defined. 
This may be a reflection of global versus local Cr vacancies in these two different samples, as local concentrations of Cr vacancies would have an analogously local impact on $\vec{D}_{ij}$, and should not significantly alter the magnitude of $\vec{D}_{ij}$ throughout the bulk of the crystal.

Figure \ref{fig:LTEM-fast} also reveals that the soliton lattice of the short-period crystal contains various ``defects'' in the spin texture.\supercite{paterson_order_2019} 
These include shearing, where solitons are observed to bend across tens of nanometers along $c$ (marked in Figure \ref{fig:LTEM-fast} with green arrows), and dislocations, where soliton walls suddenly terminate in the $ab$ plane (marked in Figure \ref{fig:LTEM-fast} with blue arrows).
Based on the atomic resolution HAADF-STEM imaging in Figure \ref{fig:Nd-HAADF}, we propose that these spin texture defects of shearing and dislocations, which are observed in the short-period sample (but are notably absent in the long-period sample), may arise or nucleate at regions of clustered Cr vacancies which form corresponding defects in the intercalant (magnetic) sublattice.
Moreover, the presence of both dark and light stripes in Figure \ref{fig:LTEM-fast}b-e reveals a mix of left- and right-handed solitons which appear with opposite contrast under identical underfocussed imaging conditions, demonstrating heterochirality of magnetic order in the short-period sample, which is also not observed in the long-period sample.

These stark differences in the homogeneity of the magnetic order between the short- and long-period samples is again consistent with their relative cooling rates after crystal growth. 
Faster cooling rates result in pockets of locally fully occupied and ideally ordered lattices with strong DM interactions within given helical magnetic domains, but more global lattice disorder which in turn give rise to heterochirality in the magnetic lattice and CSL winding. 
Slower cooling rates result in more homogeneously distributed disorder/vacancies which uniformly weaken DM interactions and lead to a low density of helical domains, but preserve homochiral heligmagnetism.

At 315 mT (the largest field we can reach with the lens configuration used for LTEM imaging) the sample appears nearly transitioned to the FFM phase, although some soliton and domain walls remain visible (Figure \ref{fig:LTEM-fast}f). 
As with the long-period sample, the critical field \Hc\ extracted from cryo-LTEM experiments ($\>$315 mT) is larger than that determined from the bulk magnetization plateau ($\sim$200 mT). 
The residual domains appear to correlate to some degree with regions of the sample that hosted either left- or right-handed soliton walls, pointing to the inherent handedness of these regions within the crystal.
The lattice-scale configuration of spin texture across one of these left-to-right hand domain walls has not been quantitatively determined, but could involve a ``trapped'' whole or fractional soliton winding at the domain boundary.\supercite{togawa2015magnetic} 
The sizes of these single-handed domains likely also impact the discrete evolution of $L(H)$ through soliton confinement effects. \supercite{wilson2013discrete,kishine2014topological,togawa2015magnetic}
Similar to the long-period sample, the magnetic evolution in the short-period \CrxNbS\ also shows significant hysteresis as the external field is ramped up or down (Supplemental Figure S9) as well as abrupt, step-wise changes in local soliton spacing. 
Indeed, this sample demonstrates that the transitions can be extremely abrupt. In some areas we observe a $\sim 5\times$ change in local $L(H)$ as the applied field is decreased within the window of 1 mT (Supplemental Figure S10).

\section*{Conclusion}

The stark differences in the static zero-field magnetic order and the dynamic evolution of chiral magnetic textures in two ostensibly chemically identical specimens of \CrxNbS\ emphasize both the promise of these materials as highly tunable magnetic systems, as well as the pressing need to understand how nanoscale disorder in various manifestations impacts the meso- and bulk-scale properties. 
Our results show the large degree to which chiral helimagnetic textures in \CrxNbS\ (and likely extended to analogous materials such as \CrxTaS\ and \CoxMS)\supercite{du2021topological, takagi_spontaneous_2023, parkin_magnetic_1983} depend strongly on the subtle variations in single crystal synthesis, and the varied effects of structural lattice disorder within \CrxNbS\ on spin texture disorder, such as discrete jumps between different states, discontinuous evolution in $L(H)$, and the coexistence of heterochirality.
Our findings also shed light on, and may offer some explanations for, discrepancies within the literature, in which the properties of \CrxNbS\ vary across different studies.\supercite{togawa2012chiral, mito_geometrical_2018, hall_comparative_2022, ogloblichev_magnetic_2017} 
Future work should leverage additional approaches, including theoretical methods, to systematically and explicitly investigate how lattice structure, exchange interactions, and vacancy distribution -- especially in such dilute limits -- lead to the observed behaviors. 
Here, we begin to bridge the gap across these multiscale hierarchies by demonstrating how different distributions of Cr vacancies can lead to global changes in the magnetic properties.
By directly connecting bulk magnetization with the real-space evolution of the helical ordering in these Cr-intercalated materials and the atomic-scale origins of these behaviors, our work paves the way for future engineering of functional devices based on this rich class of materials.

\section*{Methods}

\subsection*{Single crystal growth}
Single crystals of \CrxNbS\ and \CrxTaS\ were grown using chemical vapor transport. Powders of elemental Cr ($-100$ $+325$ mesh, 99.97\%, Alfa Aesar), Nb ($-325$ mesh, 99.99\% excluding Ta, Ta $\leq$ 500 ppm, Alfa Aesar), Ta ($-100$ mesh, 99.98\% metals basis, Nb 50 ppm, Alfa Aesar), S (99.999\%, Acros Organics), and \ch{I_2} (99.999\%, Spectrum Chemicals) were used as received.

\textit{Growth of long-period \CrxNbS\ sample}

Elemental Cr (22.5 mg, 0.48 equiv.), Nb (84.2 mg, 1.00 equiv.), and S (60.9 mg, 2.1 equiv.) were sealed in a fused quartz ampoule (14 mm inner diameter, 1 mm wall thickness, 29 cm long) under vacuum (approximately $1 \times 10^{-5}$ Torr), along with 90.1 mg \ch{I_2} (2 mg/cm$^3$). The ampoule was placed in a MTI OTF-1200X-II two-zone tube furnace with the hot zone maintained at 1100 °C and the cold (growth) zone maintained at 1000 °C for 14 days, before cooling to room temperature at 20 °C/hour. Plate-shaped crystals with a hexagonal habit up to approximately $2 \times 2 \times 0.5$ mm were obtained.\bigskip\

\textit{Growth of short-period \CrxNbS\ sample}

Elemental Cr (25.7 mg, 0.35 equiv.), Nb (131.9 mg, 1.00 equiv.), and S (90.6 mg, 2.00 equiv.) were sealed in a fused quartz ampoule (14 mm inner diameter, 1 mm wall thickness, 48 cm long) under vacuum (approximately $1 \times 10^{-5}$ Torr), along with 38.5 mg \ch{I_2} (1.3 mg/cm$^3$). The ampoule was placed in a MTI OTF-1200X-II two-zone tube furnace with the hot zone maintained at 1000 °C and the cold (growth) zone maintained at 950 °C for 14 days, before cooling to room temperature at 60 °C/hour. Plate-shaped crystals with a hexagonal habit up to approximately $4 \times 4 \times 0.5$ mm were obtained.

\textit{Growth of \CrxTaS}

Elemental Cr (19 mg, 0.47 equiv.), Ta (141 mg, 1.00 equiv.), and S (53 mg, 2.1 equiv.) were sealed in a fused quartz ampoule (14 mm inner diameter, 1 mm wall thickness, 25 cm long) under vacuum (approximately $1 \times 10^{-5}$ Torr) along with 76 mg \ch{I_2} (2 mg/cm$^3$). The ampoule was placed in a MTI OTF-1200X-II two-zone tube furnace, with the hot zone maintained at 1100 °C and the cold (growth) zone maintained at 1000 °C for 14 days, before cooling to room temperature at 20 °C/hour. Plate-shaped crystals with a hexagonal habit up to approximately $2 \times 2 \times 0.3$ mm were obtained.

\subsection*{Bulk characterization}

DC magnetization measurements were carried out on a Quantum Design Physical Property Measurement System Dynacool equipped with a 12 T magnet using either the Vibrating Sample Magnetometer option or the AC Measurement System II option. Single crystals were affixed to quartz sample holders with GE Varnish such that the magnetic field was applied perpendicular to the crystallographic $c$-axis. Raman spectroscopy was collected on a Horiba LabRAM HR Evolution with an ultra-low frequency filter using 633 nm laser excitation and powers between 1 and 8 mW. Energy dispersive X-ray spectroscopy was acquired on a FEI Quanta 3D FEG or a Scios 2 DualBeam scanning electron microscope with an accelerating voltage of 20 kV.

\subsection*{Electron microscopy}
Cross-sectional S/TEM samples were prepared by the standard focussed ion beam (FIB) lift-out method in Thermo Fisher Scientific Helios G4 UX or Scios 2 DualBeam. 
Samples were thinned to maximize the total area of electron transparency ($t/\lambda \sim 1-1.5$ at 300 kV = $\sim 90-135$ nm) for magnetic imaging (Supplemental Figure S6). 
High-angle annular dark-field (HAADF)-STEM images were collected in the thinnest parts of the prepared lamellae, near the top edge, in regions which were significantly thinner ($t/\lambda \sim 0.3$ at 300 kV = $\sim 30$ nm). 
The intercalant (Cr) order was found to be extremely sensitive to the high-energy STEM probe, so HAADF-STEM imaging was performed on a Thermo Fischer Spectra 300 X-CEFG operating at 120 kV with a low probe currents $<$ 20 pA, convergence angle of 24 mrad, and inner (outer) collection angles of 68 (200) mrad. 
To minimize the possibility of introducing additional disorder in the intercalant lattice during measurement, all HAADF-STEM images were acquired from regions without prior exposure (other than limited doses at very low magnification for the purposes of sample alignment).
For each image, a small (few-nm$^2$) nearby region was used to focus the STEM probe before blanking the beam to navigate to a fresh region of the specimen and letting the stage stabilize for to atomic-resolution image acquisition. 
In this way, large total areas of each sample were imaged with atomic-resolution to build a statistical picture of the relative density of clustered Cr vacancies. 

Cryogenic Lorentz transmission electron microscopy (cryo-LTEM) was performed in an FEI Titan Themis operating at 300 kV.
A Gatan 636 side-entry double-tilt liquid nitrogen holder was used to cool the samples to $\sim$100 K. 
Both \CrxNbS\ samples were cooled simultaneously in the TEM to eliminate the possibility of additional variation in magnetic textures due to cooling rate during the experiment. 
The samples were cooled in zero-field conditions with a cooling rate on the order of -0.05 K/s.
Over-, under-, and in-focus LTEM images were collected on a Ceta CCD camera with 0.5-4 sec. acquisition times. 
Variable magnetic field was applied by adjusting the strength of the of the objective lens according to previous calibration of the system. 
To correct for the image shift and rotation caused by changing fields during \emph{in situ} data collection, each frame in the LTEM series was realigned by rotation and rigid shift. 
We ensured that the magnetic order was robust to the TEM electron radiation through progressive exposure series for each sample and found no changes under the conditions used for this study.

\section*{Acknowledgements}
The authors thank Ariana Ray for useful discussions. 
This material is based upon work supported by the Air Force Office of Scientific Research under AFOSR Award No. FA9550-20-1-0007. B.H.G. was supported by the University of California Presidential Postdoctoral Fellowship Program (UC PPFP) and by Schmidt Science Fellows in partnership with the Rhodes Trust. O.G. acknowledges support from an NSF Graduate Research Fellowship grant DGE 1752814, and National GEM Consortium Fellowship. L.S.X. acknowledges support from the Arnold and Mabel Beckman Foundation (award no. 51532) and L'Or\'{e}al USA (award no. 52025) for postdoctoral fellowships. Confocal Raman spectroscopy was supported by a Defense University Research Instrumentation Program grant through the Office of Naval Research under award no. N00014-20-1-2599 (D.K.B.). Electron microscopy was supported by the Platform for the Accelerated Realization, Analysis, and Discovery of Interface Materials (PARADIM) under NSF Cooperative Agreement No. DMR-2039380. This work made use of the Cornell Center for Materials Research (CCMR) Shared Facilities, which are supported through the NSF MRSEC Program (No. DMR-1719875). The FEI Titan Themis 300 was acquired through No. NSF-MRI-1429155, with additional support from Cornell University, the Weill Institute, and the Kavli Institute at Cornell. The Thermo Fisher Helios G4 UX FIB was acquired with support by NSF No. DMR-1539918. The Thermo Fisher Spectra 300 X-CFEG was acquired with support from PARADIM (NSF MIP DMR-2039380) and Cornell University. Other instrumentation used in this work was supported by grants from the Canadian Institute for Advanced Research (CIFAR–Azrieli Global Scholar, Award no. GS21-011), the Gordon and Betty Moore Foundation EPiQS Initiative (Award no. 10637), the W.M. Keck Foundation (Award no. 993922), and the 3M Foundation through the 3M Non-Tenured Faculty Award (no. 67507585).

\section*{Competing Interests}
The authors declare no competing interests.

\printbibliography
\end{document}